\documentclass{aa}  

\usepackage{graphicx}
\usepackage{txfonts}
 \usepackage{xcolor}
\begin{document}

   \title{A possible mechanism for multiple changing look phenomenon in Active Galactic Nuclei}


   \author{M. Sniegowska
          \inst{1,2}          \and
          B. Czerny\thanks{bcz@cft.edu.pl}\inst{2} \and
          E. Bon\inst{3} \and
          N. Bon\inst{3}           }

   \institute{Nicolaus Copernicus Astronomical Center (PAN), ul. Bartycka 18, 00-716 Warsaw, Poland\\
         \and Center for Theoretical Physics, Polish Academy of Sciences, Al. Lotnik\'ow 32/46, 02-668 Warsaw, Poland\\
         \and Astronomical Observatory  Belgrade, Volgina 7,  11060 Belgrade, 
Serbia   }


 
  \abstract
   {Changing-look phenomenon observed now in a growing number of active galaxies challenges our understanding of the accretion process close to a black hole.  }
   {We propose a simple explanation for the sources where multiple semi-periodic outbursts are observed, and the sources are operating at a few per cent of the Eddington limit.}
   {The outburst are caused by the radiation pressure instability operating in the narrow ring between the standard gas-dominated outer disk and the hot optically thin inner Advection-Dominated Accretion Flow. The corresponding limit cycle is responsible for periodic outbursts, and the timescales are much shorter than the standard viscous timescale due to the narrowness of the unstable radial zone.
   }
   {Our toy model gives quantitative predictions and works well for multiple outbursts like those observed in NGC 1566, NGC 4151, NGC 5548 and  GSN 069, although the shapes of the outbursts are not yet well modeled, and further development of the model is necessary.}
   {}


   \maketitle
%

\section{Introduction}
\label{sec:intro}

Active Galactic Nuclei have been always known as strongly variable sources in most of their broad band spectra (e.g. IR: \citealt{edelson1987,kozlowski2016}; optical: \citealt{ulrich1997,kawaguchi1998,sesar2007}; X-ray: \citealt{ulrich1997,lawrence1993}). Most of the variability can be attributed to variations of the red noise character, both in the optical and in the X-ray band \citep{mchardy1987,lehto1993,czerny1999,gaskell2003}. However, some of the observed changes lead to far more dramatic changes than expected from the red noise trend. These changes sometimes are revealed  in the temporary change of the source classification, and these sources started to be known as Changing-Look AGN (CL AGN; \citealt{matt2003}).
There is no well established definition what can - or cannot - be classified as a CL AGN, and we adopted the view that the name can be used for the broad class of objects, not necessarily showing confirmed changes in the optical flux. With progressing understanding of the mechanisms, proper classification will be certainly introduced.

CL AGN phenomenon was once considered as rather rare.  The changes corresponded either to a drastic change in X-ray spectrum, or in the optical/UV emission lines and continuum, depending on the studied wavelength range
\citep{bianchi2005,denney2014,Shappee2014}. On the other hand, historical lightcurves of nearby sources, including well studies AGN  \citep[e.g.][]{cohen1986,iijima1992,storchibergmann1993,bon2016,oknyanski2016,shapovalova2019} indicated that such episodes do happen. With more and more optical and X-ray surveys, the number of CL AGN is rapidly growing \citep{ruan2016,ross2018,yang2018,stern2018,trakhtenbrot2019,macleod2019}, and the question about the mechanism of the phenomenon must be addressed. The most extreme case of such phenomenon in the form of Quasi-Periodic Eruptions QPE) has been recently discovered by \citet{Miniutti2019} and \citet{2020giustini}.

There is still an on-going discussion whether the phenomenon is intrinsic to the central engine of the active galaxy, or it is just a result of a temporary obscuration or disappearance of such obscuration. While for some CL AGN phenomenon the obscuration mechanism can work, for most of the sources there are strong arguments in favor of the intrinsic changes:

 $\bullet$ complex multi-band recovery, inconsistent with obscuration \citep[e.g.][]{mathur2018}
 
  $\bullet$ strong changes seen in the IR, where the obscuration should not play a role \citep{sheng2017,stern2018}
  
 $\bullet$ low level of polarization in CL AGN which argues against the scattering (and obscuration) scenario \citep{2019arXiv190403914H}
 
 $\bullet$ different variability behaviors of the observed emission lines in spectra of CL AGN \citep[e.g.][]{kynoch2019}
 
 $\bullet$ regular QPE behavior cannot be due obscuration because of the characteristic spectral evolution during outbursts \citep{Miniutti2019,2020giustini}. 
 
 Thus in most sources the intrinsic change in the bolometric luminosity later affects the X-ray and broad line region (BLR) appearance. These intrinsic changes can be either related to Tidal Disruption Event (TDE), or be a result of the spontaneous unforced behavior of the accretion flow close to a black hole. In some cases perhaps TDE provides the answer but in sources with repeated events the TDE is statistically unlikely. 

In the present paper we concentrate on the discussion of the plausible mechanism which can lead to regular or semi-regular repeating outbursts intrinsic to the nucleus. In such case the source behavior should be related to some instabilities in the accretion flow. However, the radiation pressure instability expected to be operational in the innermost part of an AGN accretion disk does not provide the proper timescales \citep[e.g.][]{gezari2017}. Convenient formulae for the duration of such outbursts given in \citep{grzedzielski2017} give timescales of hundreds of years for a black hole mass of $10^7 M_{\odot}$. \citet{dexter2019} suggested that strong magnetization can shorten the estimated timescales. On the other hand, we can look for another mechanism related to the complexity of the innermost part of the flow, and \citet{noda2018} proposed that the CL behavior in the source Mrk 1018 is related to the temporary disappearance of the warm corona. The source NGC 1566 notable for numerous CL outbursts \citep[e.g.][]{alloin1986, baribaud1992, oknyansky2019} does not show the presence of the warm corona component before the outburst \citep{parker2019}. The present observations cannot resolve directly any of these issues since they show at best the presence of the gas reservoir at a distance of 60 pc from the black hole (Mkn 590; \citealt{raimundo2019}). They only show that the phenomenon is complex, for example the reappearance of broad lines in Mkn 590 is not accompanied by the full recovery of the continuum \citep{raimundo2019}.

In this paper we propose a new mechanism which is suitable for explaining regular outbursts in sources which are not very close to the Eddington ratio. Using highly simplified toy model we aim at discussion whether the mechanism is likely to reproduce the observed timescales and therefore deserves the effort of more detailed description in the future.

\section{Analytical estimates and the model geometry}
\label{sec:analitical}

The character of the accretion flow in AGN strongly depends on the Eddington ratio of the source. In sources with the Eddington ratio above a few per cent, optically thick, geometrically thin disk extends down to the Innermost Stable Circular Orbit (ISCO). Modelling of the optical/UV emission of quasars support this view \citep[e.g.][]{capellupo2015}, although warm corona seems to be needed to explain the soft X-ray excess. However, low luminosity AGN, showing low-ionization nuclear emission line region (LINERS) do not show such a component in the optical/UV spectra, and it is generally accepted that in these sources the innermost part of the accretion flow proceeds in a form of an optically thin advection-dominated accretion flow (ADAF). 

For simplicity, we introduce here a definition of the Eddington accretion rate based on Newtonian physics:
\begin{equation}
\dot M_{Edd} = {48\pi GM_{BH} m_p \over \sigma_T c},
\end{equation}
where $M_{BH}$ is the black hole mass, $m_p$ is the proton mass, and $\sigma_T$ is the Thomson cross-section. We thus measure the ratio of the accretion rate to the Eddington accretion rate using $\dot m = \dot M /\dot M_{Edd}$.

In those units, the transition between an inner ADAF flow and an outer standard accretion disk \citep{abramowicz1995,czerny2019} takes place at
\begin{equation}
R_{ADAF} = 2 \alpha_{0.1}^4 \dot m^{-2} R_{Schw},
\end{equation}
where $\alpha_{0.1}$ is the viscosity parameter introduced by \citet{ss73}, in units of 0.1, and $R_{Schw} = 2 GM_{BH}/c^2$ is the Schwarzschild radius of the black hole.

Standard accretion disk is unstable in the innermost part, when the radiation pressure dominates \citep{le74,pringle1973,ss76}, and the transition from the outer stable to the inner unstable radius takes place at: 
\begin{equation}
R_{tr} = 1522 (\alpha_{0.1}m_7)^{2/21} \dot m^{16/21} R_{Schw},   
\end{equation}
\citep{ss73}. Here $m$ is the black hole mass expressed in units of $10^7 M_{\odot}$.  
The two lines cross at the specific accretion rate, $\dot m_{st}$, 
\begin{equation}
\label{eq:limit}
\dot m_{st} = 0.0905 \alpha_{0.1}^{41/29} m_7^{-1/29},
\end{equation}
where the dependence on the black hole mass is negligible, but the dependence on the viscosity coefficient is stronger than linear. The radius where it happens is given by:
\begin{equation}
R_{st} = 244 \alpha_{0.1}^{43/29}  m_7^{2/29} R_{Schw}.
\end{equation}

If the accretion rate of the flow is smaller than this limiting value, $\dot m_{st}$, the whole flow is stable since both the gas-dominated cold outer disk and the inner ADAF flow are stable solutions. On the other hand, if the accretion rate is higher that $\dot m_{st}$, there is a disk range, dominated by the radiation pressure which is unstable and could lead to a limit cycle behavior.

\begin{figure}
\centering
\includegraphics[width=0.45\textwidth]{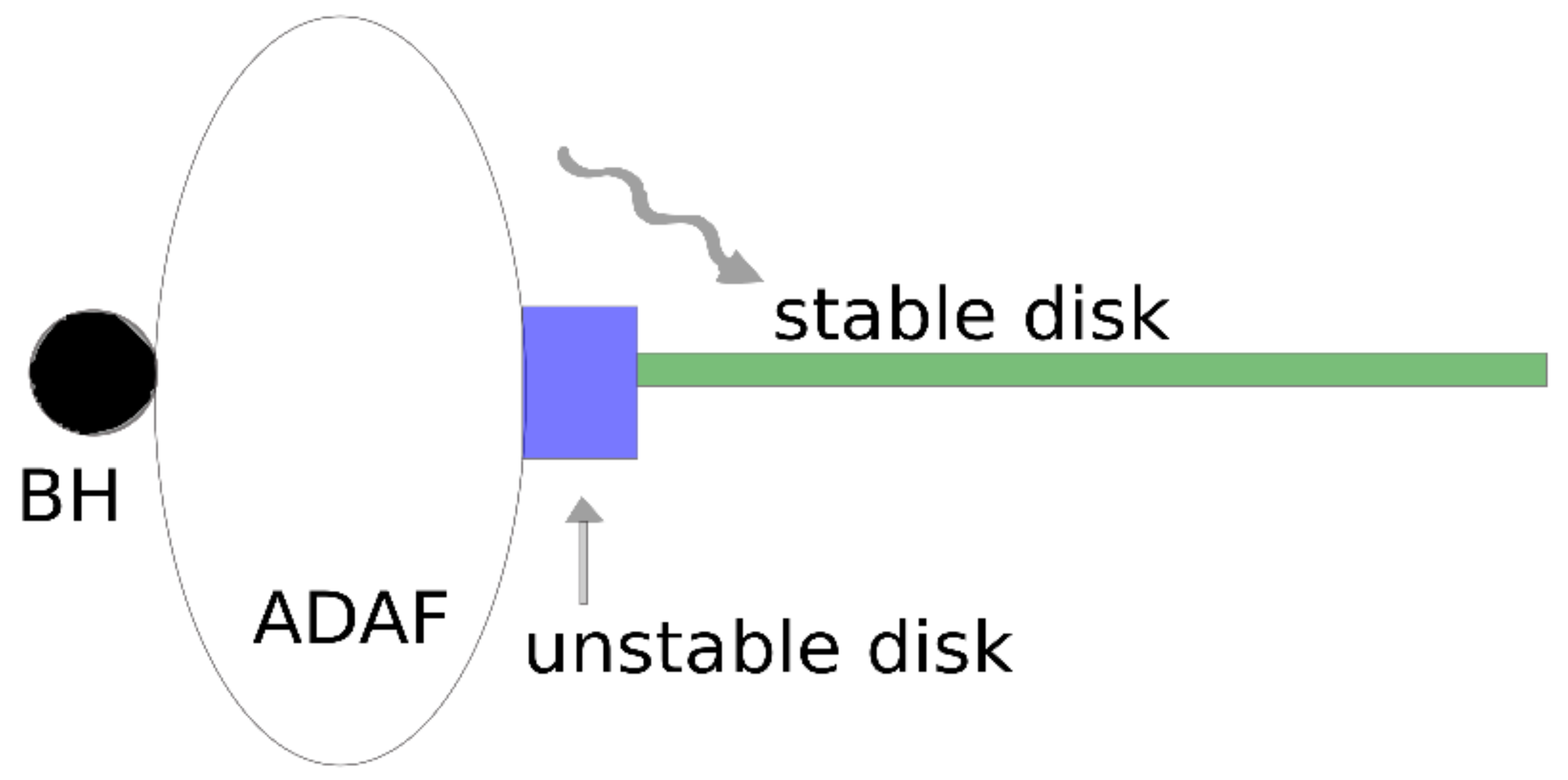}
\caption{The schematic view of the innermost part of the flow: outer cold stable disk (green), intermediate zone (disk part unstable due to radiation pressure instability), and inner hot ADAF which illuminates the outer disk.}
 \label{fig:schematic}
\end{figure}

The timescale for such oscillations is generally set by the viscous timescale of the Shakura-Sunyaev disk, 
\begin{equation}
\tau_{visc,SS} = {1 \over \alpha} ({R \over H})^2 ( {R^3 \over G M_{BH}})^{1/2}
\end{equation}
(\citealt{ss73}; see e.g. a review by \citealt{czerny2006}) 
which is long for the case of AGN, such as hundreds of years. However, if $\dot m$ is just above the treshold defined by Equation~\ref{eq:limit}, then the radial extension of the unstable zone, $\delta R$ is much smaller than the radius $R$ itself. We illustrate schematically such a geometry in Figure~\ref{fig:schematic}. In that case, the time needed to empty the zone is reduced, and the viscous evolution will happen in a timescale
\begin{equation}
\tau_{visc} = \tau_{visc,SS} {\Delta R \over R}.
\end{equation}

Thus, for sources at low Eddington accretion rate the radiation pressure instabillity, operating in a very narrow zone at the border between the outer standard disk and an inner ADAF flow can provide a viable mechanisms explaining repeating outbursts in some CL AGN in timescales of a few years. The schematic view of our new model of CL AGN is shown in Figure~\ref{fig:schematic}.

The small radial extension of the instability zone reduces also the amount of variable radiation flux by the same factor. On the other hand, the zone regulates the accretion flow in the inner ADAF, and most of the radiation is actually produced there, in the form of X-rays. The efficiency of the inner ADAF flow is not well known, but significant part of the energy goes directly to electrons, and subsequently, to radiation \citep[e.g.][]{bisnovatyi1997,yuan2014,marcel2018}. The outburst thus should be clearly seen in X-rays, but in addition X-ray irradiation of the cold disk will lead to enhancement of the disk emission in optical/UV band.

\section{One-zone time-dependent toy model}
\label{sect:toy_model}

We construct a simple toy model in order to check whether the mechanism may indeed give repeated outbursts of the observationally required properties. We basically follow the 1-D model of time evolution of accretion disk under radiation pressure instability \citet{janiuk2002}, but we simplify it further by concentrating on a single zone approximation, thus reducing the numerical problem to ordinary differential equation in time. Instead of solving for vertically averaged disk structure as functions of both time and disk radius, we follow the time-dependent evolution of a single zone, representing the radiation pressure dominated region. It is a good approximation if the radial extension of the zone is small, i.e. the zone is narrow and comparable to the disk thickness.

The evolution of the zone in thermal timescale is very similar to the one in multi-radius approach, as the thermal evolution results from the net effect of the heating, radiative cooling, and advection cooling. We assume that the disk is in hydrostatic equilibrium.
However, the viscous evolution is now set simply by the boundary conditions: (i) between the zone and outer stationary disk (ii) between the zone and inner ADAF:
\begin{itemize}
\item at the border between the zone and the outer disk, we assume a constant inflow rate of the material provided by the stable outer disk, $\dot M_0$. 
\item at the border between the zone and the inner ADAF, the material is removed from the disk at a rate $\dot M$ by evaporation due to the electron conduction. 
\end{itemize}

Since this is a simple toy model, we do not use any advanced description of this process which would require the knowledge of the ADAF density, ion and electron temperature \citep[e.g.][]{rozanska00,2007liu,2020qiao}. Instead, we postulate that the efficiency of the process should be proportional to the zone height, since the hot inner ADAF is geometrically thick so the interacting surface is set by the cold disk state. We additionally assume that evaporation is more efficient when there is more mass in the unstable zone. If the zone and ADAF happen to be in equilibrium, then the outflow rate from the zone should be equal to the inflow rate. So introducing the equilibrium zone thickness $H_0$, and equilibrium surface density in the disk, $\Sigma_0$ our approach allows us to specify the evaporation rate in general as
\begin{equation}
\dot M = \dot M_0 {H \over H_0} {\Sigma \over \Sigma_0},
\label{eq:evap}
\end{equation}
where the quantities $H$ and $\Sigma$ describe the height and surface density of the evolving zone. We always start our time-dependent evolution from an equilibrium model, but if the solution corresponds to unstable one, the disk will perform the limit cycle.

From the two assumptions above, we obtain the time evolution of the surface density in the zone
\begin{equation}
\label{eq:sigma}
{d \Sigma \over dt} = {\dot M_0 - \dot M \over 2 \pi R \Delta R,}
\end{equation}
i.e. it is given by the imbalance between a constant inflow rate into the zone, $\dot M_0$, and variable outflow rate $\dot M$ from the zone to inner ADAF flow. 
  
The evolution of the zone in the thermal timescale, given by Equation 33 in \citet{janiuk2002}, under our assumptions, for a narrow zone, reduces to the following equation for the equatorial disk temperature, $T$: 
\begin{eqnarray}
\label{eq:temp}
{d \log T \over dt}& = &{(Q^+ - Q^{-} -Q_{adv})(1 + \beta) \over P H [(12 - 10.5 \beta)(1 + \beta) + (4 - 3\beta)^2]} \nonumber \\
&+&2 {d \log \Sigma \over dt}{4 - 3\beta \over (12-10.5 \beta)(1+\beta) + (4 - 3\beta)^2}.
\end{eqnarray}
Here the calculation of the derivatives of the disk thickness $H$ are already included in the expression. The values of the disk thickness, total pressure $P$, gas to the total pressure ratio, $\beta$, viscous heating $Q^{+}$, radiative cooling $Q^{-}$  are determined from the standard equations of the vertically averaged disk structure in hydrostatic equilibrium as in \citet{janiuk2002}, but here we do not introduce any additional correction coefficients related to the disk vertical structure (like $C_1$, $C_2$) since the current model is very simple. The advection cooling term $Q_{adv}$ is determined as
\begin{equation}
\label{eq:adv}
 Q_{adv} = {\dot M P H \over 2 \pi R \Delta R \Sigma},   
\end{equation}
so we include only advection term related to the inflow from the zone to inner ADAF, and we neglect the energy carried into the zone from the outer disk, which should be negligible.

Thus time-dependent partial differential equations (26) and (33) from \citet{janiuk2002} reduce to ordinary differential equation for the time evolution of a surface density and temperature in the equatorial plane of a single zone.

The geometrically narrow instability zone evolves fast, but the amount of energy dissipated in this zone is also correspondingly small. Therefore, the changes in the zone luminosity by itself does not change significantly the system luminosity. However, the zone acts as a regulator of the accretion flow in the innermost ADAF. 

ADAF flow was frequently considered as inefficient, but most estimates of the ion-electron coupling and of the Ohmic heating imply that actually ADAF flow in energetically quite efficient, at least when the accretion rate is not many orders of magnitude below the Eddington accretion rate. 
\citep[e.g.][]{bisnovatyi1997, 2011ferreira, 2018Galax...6..122H}. Therefore, the inner part of the flow generates more energy than the outer part of the disk and the transition zone (the exact number would depend on the black hole spin). This energy is emitted in X-rays but part of the produced X-ray radiation will illuminate the disk and enhance the disk emission. 

We thus assume the typical flow efficiency of 10\%  in ADAF and calculate the result of the disk irradiation. ADAF is an extended medium so in principle this is a complex 2-D issue but in our simple model we represent the ADAF emission by emission localized along the symmetry axes since that allows us to calculate the effect in a simple way (we used the method and the code developed in \citet{loska2004}. This irradiation is very important, strong illumination is observed in reverberation-studied sources like NGC 5548  \citet{2015edelson, 2015ApJ...806..128D, 2016ApJ...824...11G,2017edelson, 2016ApJ...821...56F, 2017ApJ...835...65S, 2018mchardy, 2019kriss} where the variable X-ray emission drives the accretion disk continuum variability, although the correlation is not always perfect. 

In our toy model we assume, for simplicity, that the inner region luminosity is equal to the total (time-dependent) bolometric luminosity
\begin{equation}
L_{ADAF} = \eta \dot M c^2,
\end{equation}
with the flow efficiency $\eta$ equal 0.1 like in radiatively efficient flow. The illumination of the outer disk is calculated semi-analytically as in \citet{loska2004}, assuming that the emission is localized along the symmetry axis; otherwise 3-D computations would be necessary. The emissivity is adopted as a power law with index $\beta = 2.0$, the maximum distance is equal to the transition radius $R_{tr}$, and the minimum distance along the axis is set at 1/3 of this value. We assume complete local thermalization of the incident flux by the cold disk. 

In principle, free parameters of our models are the black hole mass, the accretion rate and the viscosity as free parameters, since the location of the transition and the extension of the unstable zone should results from the computations of the disk structure. However, our toy model does not have all the ingredients (like proper description of the vertical structure, opacity, convection, irradiation etc., see for example \citealt{rozanska1999}). So we additionally treat the radius and the zone width as independent parameters.

\section{Results}
\label{sect:results}

We use now our toy model of radiation pressure instability in a narrow zone between the outer cold disk and an inner hot flow to model the repeating outbursts observed in some AGN. The model parameters are: the external accretion rate, $\dot M_0$, the radius, $R$, the width of the unstable zone, $\Delta R$, and the viscosity parameter, $\alpha$, in the zone. The two remaining parameters, $\Sigma_0$ and $H_0$ are determined self-consistently from the equilibrium (unstable) solution located at the stability curve.

\subsection{Stability curve}

\begin{figure}
\centering
\includegraphics[width=0.45\textwidth]{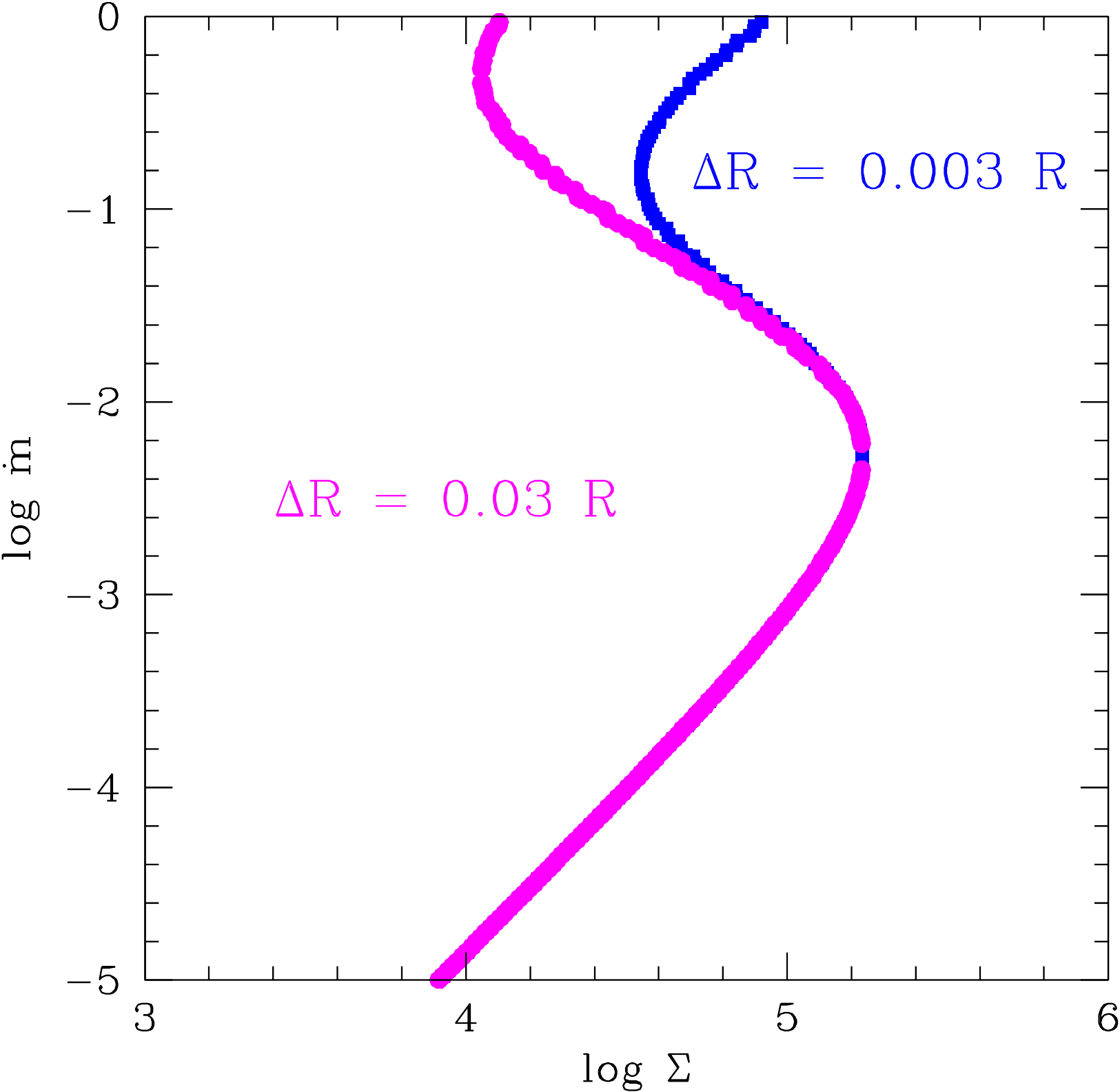}
\caption{Accretion rate vs. surface density in the transition zone between the cold SS outer disk and the inner ADAF in a stationary model in two cases $\Delta R = 0.003 R$ (blue points), and $\Delta R = 0.03 R$  (magenta points). Other parameters: log M = 6.92, $R = 30 R_{Schw}$, $\alpha = 0.02$. }
\label{fig:zone}
\end{figure}
Stability curve is built of solutions to equations \label{eq:sigma} and \label{eq:temp}, assuming that all time derivatives are equal 0. They are conveniently plotted as a function of the external accretion rate, $\dot M$. We express it in dimensionless units. In the case of the stationary solution, the accretion rate inside the zone is coupled to the zone properties as in a standard stationary disk:
\begin{equation}
\dot M = 4 \pi \alpha P H/\Omega_K,
\end{equation}
as in the standard disk of \citet{ss73}. 

 The result is shown in Figure~\ref{fig:zone} (blue line). Here we adopt parameters appropriate for NGC 1566. For black hole mass we assume $\log M = 6.92$ after \citet{woo2002}, we adopt the viscosity paramater $\alpha_{0.1} = 0.2$ (i.e. $\alpha = 0.02$) after \citet{grzedzielski2017}, and we take 30 $R_{Schw}$ for the radius. The zone width is assumed to be very narrow, 0.003 R, comparable to the disk thickness.
 
 Our stability curves in their high accretion rate parts depend on the adopted width of the zone since the advection term in our model explicitly contains it (see Equation \ref{eq:adv}).
When the zone is narrow, the advection works particularly efficiently. 

The negative slope of the stability curve implies that the solution is unstable. So, for the assumed radius and the black hole mass, the flow with accretion rate higher than $\dot m \sim 0.01$ is unstable. The upper stable branch starts quite early for a narrow zone, so the instability is expected to operate for $\dot m$ between 0.01 and 0.1 in this case. If the zone width increases, the branch stabilized by advection starts at higher accretion rates, and for $\Delta R$ of order of $R$ the stabilization would happen above the Eddington accretion rate, as in a standard slim disk \citep{abramowicz1988}. However, our toy model is not expected to work for geometrically broad zone.

\subsection{Time evolution of the accretion rate through the zone}

We compute the time evolution of the zone by assuming the value of the black hole mass, the radius, the radial width of the zone, and the viscosity parameter. We then choose the external accretion rate from the range corresponding to the unstable branch.  The disk irradiation parameters are fixed, as described in Section~\ref{sect:toy_model}.

\begin{figure}
\centering
\includegraphics[width=0.45\textwidth]{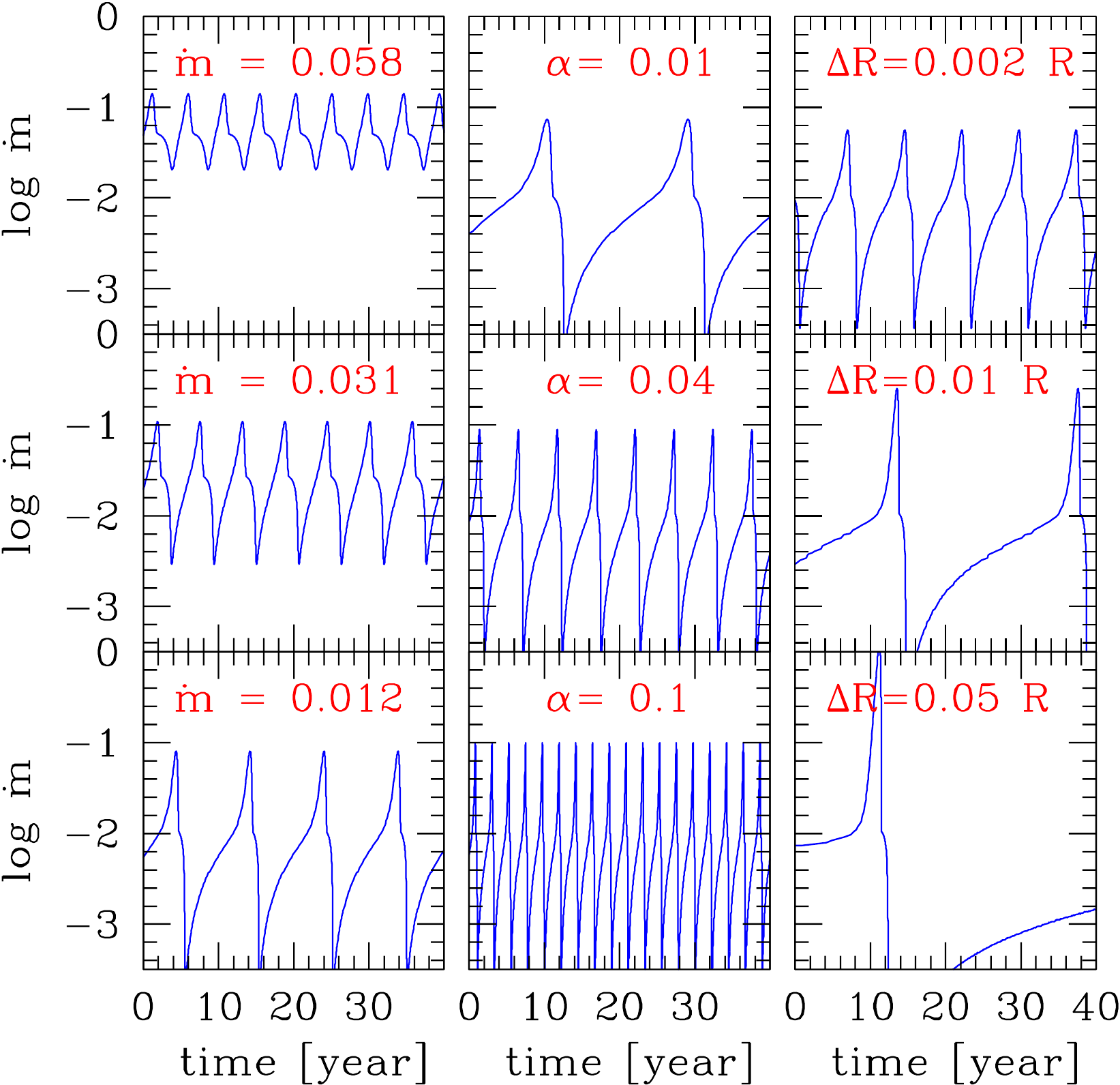}
\caption{The dependence of the time-dependent accretion rate on the external steady accretion rate $\dot m_0$, viscosity parameter $\alpha$, and the geometrical thickness of the unstable zone, $\Delta R$. Parameters are marked in each panel, the default parameters are:  $\dot m = 0.0122$, $\alpha = 0.02$, $\Delta R = 0.003 R$. Fixed parameters: black hole mass $\log M = 6.92$, inner radius of the disk $R = 30 R_{Schw}$. }
\label{fig:zestaw1}
\end{figure}

Exemplary time evolution of the accretion rate regulated by the unstable zone is given in Figure~\ref{fig:zestaw1}. We fixed the black hole mass there at the value corresponding to NGC 1566, but we varied the accretion rate, the viscosity parameters and the radial size of the unstable zone.  We see there that the timescales of outbursts, and the outbursts amplitudes are very sensitive to these parameters. The shape of the outburst vary less, and in our model the duration of the bright phase (outburst) is always longer then the outburst separation. This is because the accumulation phase is longer than the evaporation rate and the transfer through ADAF. Outbursts are regular since our toy model is very simple. More advanced models of disk instabilities, which include the wind, irradiation, magnetic field or tidal interaction with a companion in a binary system frequently lead to much more complex outbursts \citep[see e.g.][and the references therein]{hameury2019}. These, models however, do not consider the inner ADAF and narrow instability zone, so our toy model gives interesting estimates of the timescales, and further development may easily lead to more complex lightcurve shapes.

\begin{figure}
\centering
\includegraphics[width=0.45\textwidth]{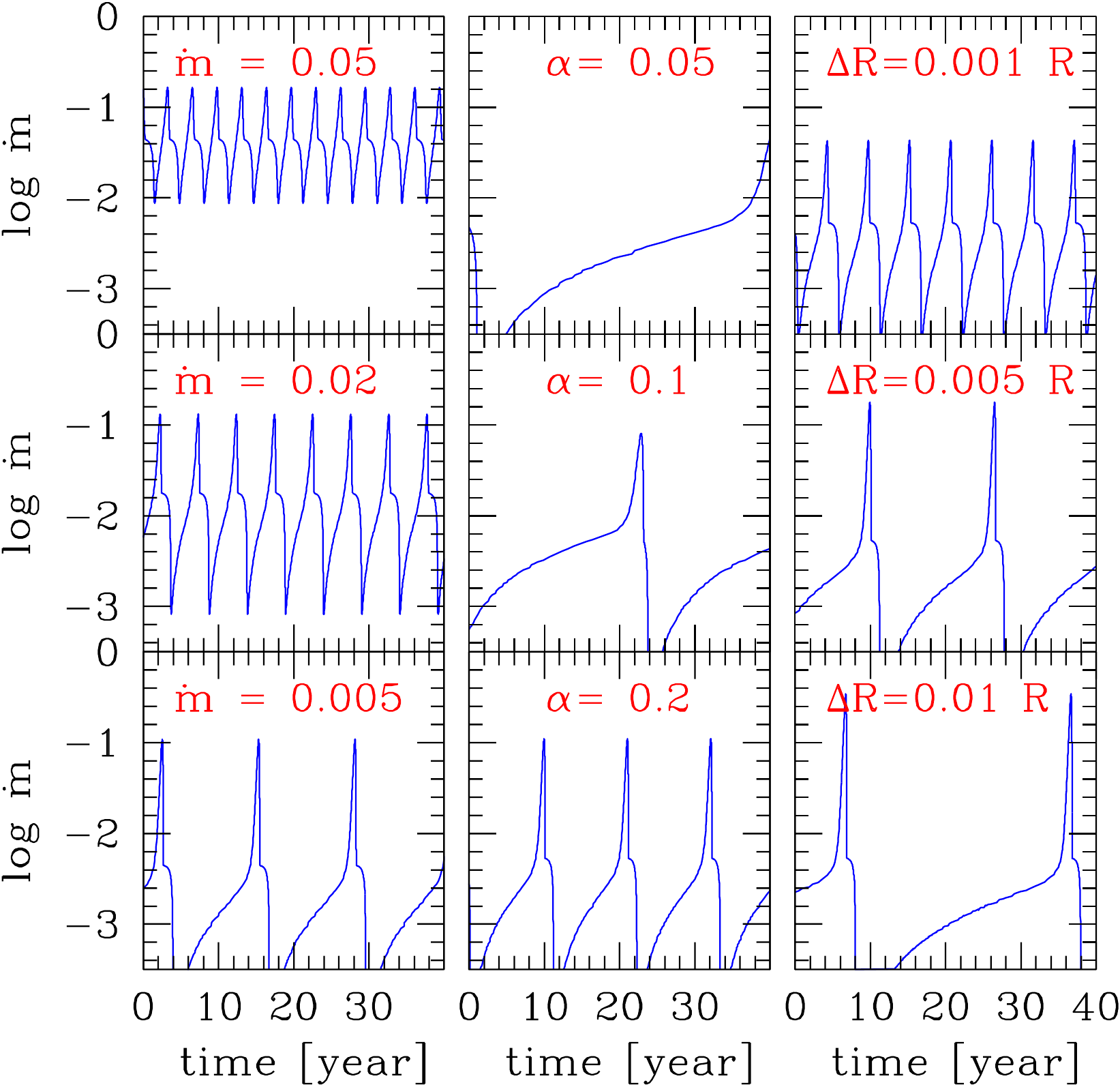}
\caption{The dependence of the time-dependent accretion rate on the external steady accretion rate $\dot m_0$, viscosity parameter $\alpha$, and the geometrical thickness of the unstable zone, $\Delta R$. Parameters are marked in each panel, the default parameters are:  $\dot m = 0.006$, $\alpha = 0.2$, $\Delta R = 0.003 R$. Fixed parameters: black hole mass $\log M = 7.94$, inner radius of the disk $R = 20 R_{Schw}$. }
\label{fig:zestaw2}
\end{figure}

The evolution is significantly slower for more massive black holes. Therefore, in Figure~\ref{fig:zestaw2} we show a set of lightcurves for a black hole mass more appropriate for sources like NGC 5548. In order to model frequent outbursts we have to request values of the higher viscosity parameter.

\subsection{Irradiation of the cold outer disk}

As discussed in Section~\ref{sect:toy_model}, the variable accretion rate in the unstable zone and in the innermost part of the flow strongly affects the outer disk.
Thus, the variable accretion rate as shown in Figure~\ref{fig:zestaw1} has to be used to receive the time evolution of the monochromatic flux and the line luminosity. 

\begin{figure}
\centering
\includegraphics[width=0.45\textwidth]{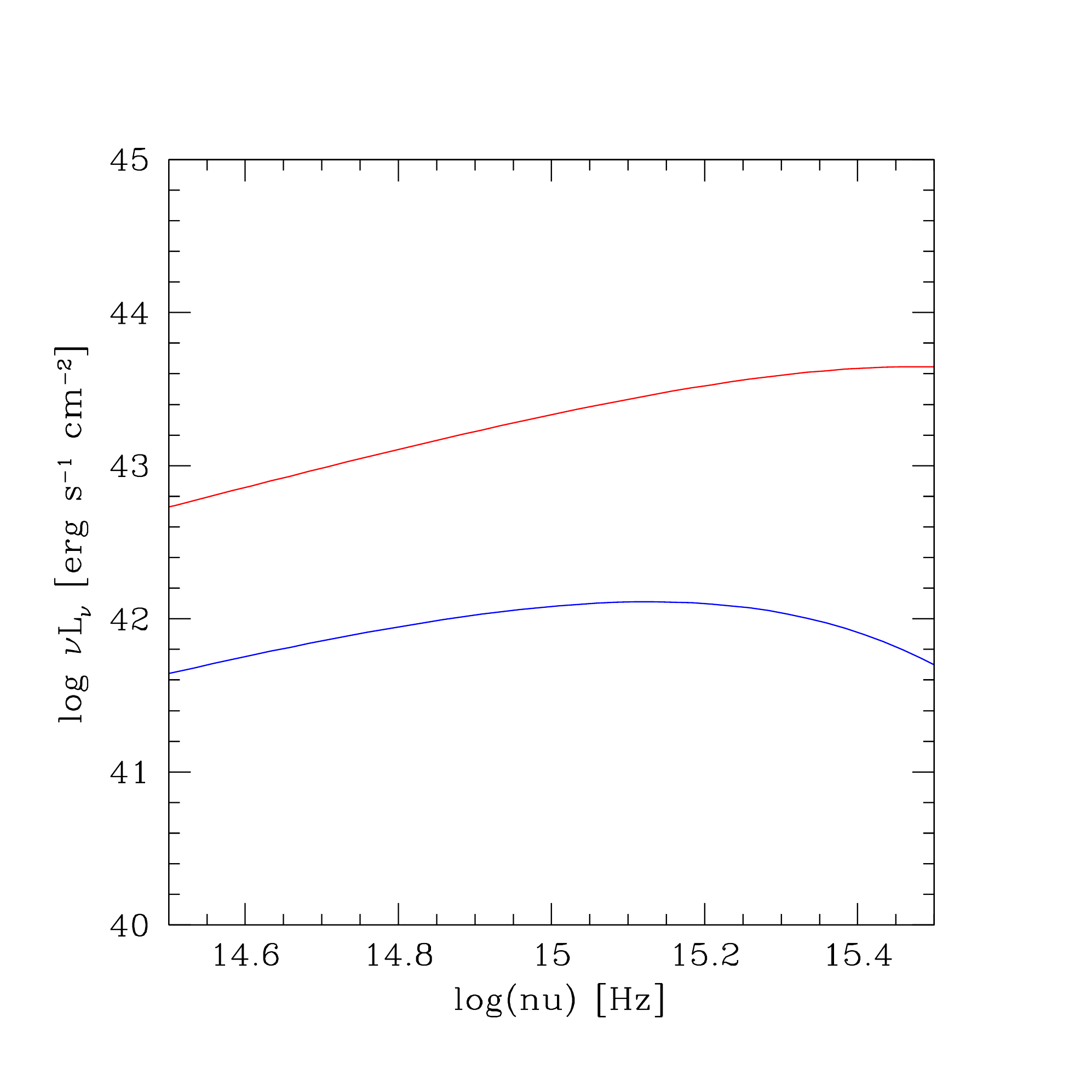}
\caption{Two extreme states of the accretion disk in the source: between outbursts (blue line) and during outburst (red line). Here we neglect the contribution from the starlight.}
\label{fig:spectra}
\end{figure}

In Figure~\ref{fig:spectra} we show two extreme examples of the spectra from an illuminated disk: between the outburst and at the peak of the outburst. For the chosen parameters, given in the figure caption, the flux at V band has changed by an order of magnitude, and the spectrum became much bluer in the far UV. Parameters which were used in this case are: $\dot m = 0.012$ ,  $\log M = 6.92$ , appropriate for NGC 1566.

\begin{figure}
\centering
\includegraphics[width=0.45\textwidth]{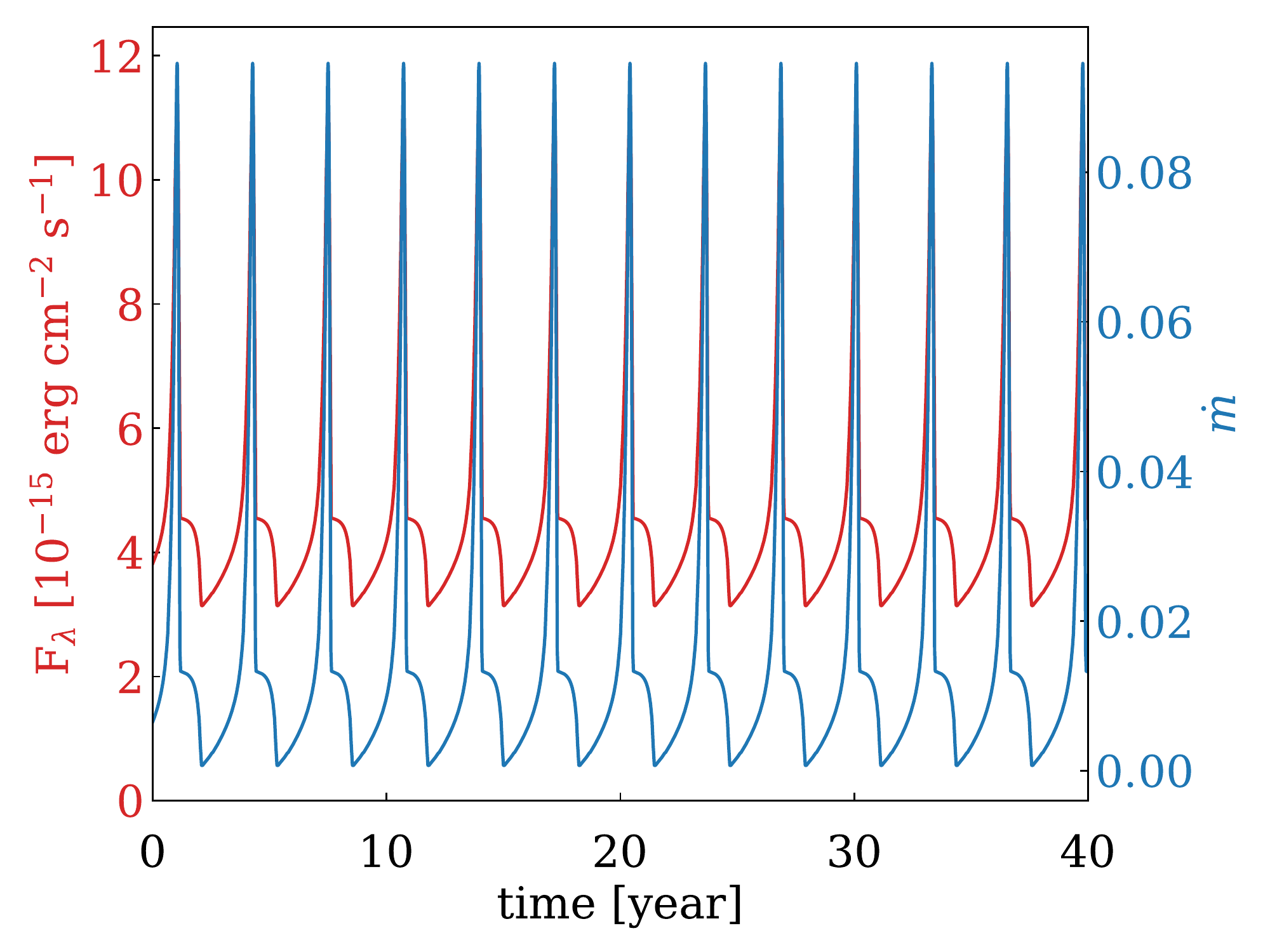}
\caption{The comparison of the modeled variations of the bolometric luminosity measured as the Eddington ration (blue line) with the modeled variations of the monochromatic disk luminosity in V band (red line), with irradiation included.}
\label{fig:mdot_V_band}
\end{figure}

We thus compare the bolometric lightcurve resulting from the instability to the corresponding monochromatic lightcurve. As explained in Section~\ref{sect:toy_model}, 
we derive the monochromatic lightcurve  taking into account the disk plus transition zone flux at V band for all the time steps of the evolution, using always the current value of $\dot M$ to calculate the disk illumination. The shape of the curve is similar, but not identical, with the shape of the accretion rate variability. An example is shown in Figure~\ref{fig:mdot_V_band}. Such lightcurve can be directly compared to the continuum lightcurve of a given source, but the observed lightcurve should be corrected for the starlight contamination.

If the line luminosity lightcurve is available, in principle we should compute the structure of the BLR but in our toy model we can assume that the line follows the bolometric luminosity of the source which is well represented by the varying accretion rate. 

\subsection{Comparison with the observational data}

Our toy model is not yet ready for detailed fitting of the observed lightcurves. What is more, such a comparison would be always inherently difficult since the observed variability in AGN is never strictly periodic. Thus our aim is to test if the model can roughly cover the characteristic variability timescales in the few exemplary sources.
Physical parameters which we obtained for each object are shown in Table \ref{table:fits-parameters}.

\subsubsection{NGC 1566}

This source, usually classified as Seyfert 1.5 galaxy (z = 0.005017 after NED\footnote{https://ned.ipac.caltech.edu/classic/}) is a well known CL AGN. Its semi-regular outbursts were already observed by \citet{alloin1986}. Later, in September 2017 the source started a spectacular brightening in optical band \citep{stanek2018}, and in X-rays \citet{parker2019}. The bolometric luminosity of the source thus strongly vary, from $\log L_{bol}= 41.4$ reported in \citet{combes2019}, up to $\log L_{bol}= 44.45$ \citep{woo2002} . \citet{parker2019} reported the Eddington ratio of 0.05 during the outburst and 0.002 between the outbursts.

\begin{figure}
\centering
\includegraphics[width=0.45\textwidth]{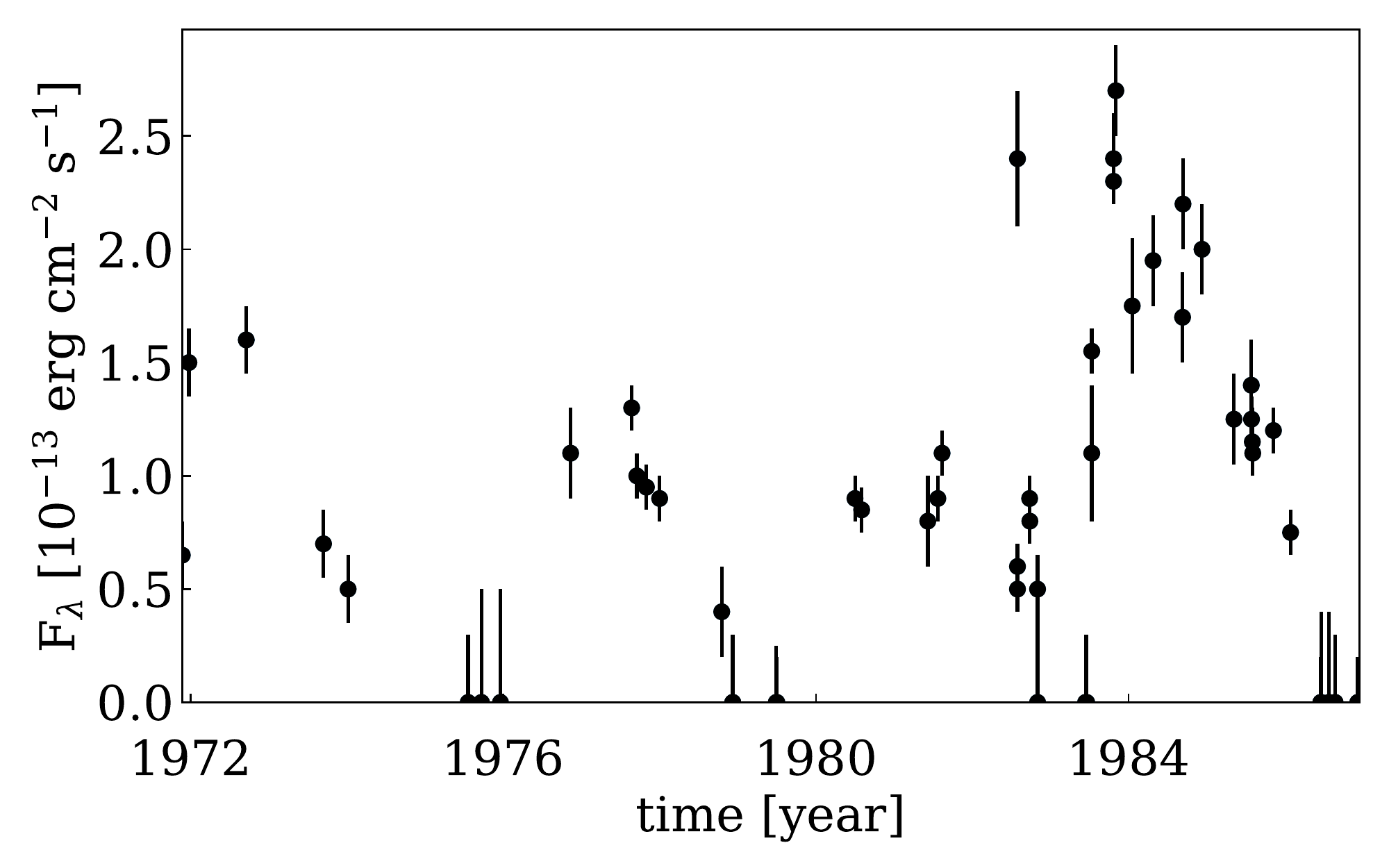}

\caption{The H$\beta$ line flux evolution in NGC 1566 from \citet{alloin1986}.}


\label{fig:V_band_curve}
\end{figure}

Mostly concentrating on the old data showing multiple outbursts (see Figure~\ref{fig:V_band_curve}) we assume that the characteristic timescale in this source is  5 years. For the black hole mass we assume the value  $\log M = 6.92$ from \citet{woo2002} (it is consistent with the value $6.8 \pm 0.3$ derived from molecular gas dynamics by \citealt{combes2019}).  We assume the mean accretion rate of $\dot m =  0.012$ in Eddington units, corresponding the mean value. 

We can find an example of the unstable solution for these input parameters assuming the value of 25 $R_{Schw}$ for the radius. The zone width is assumed to be very narrow, 0.002 R, comparable to the disk thickness. 
The required value, $\alpha = 0.04$ is by a factor 2 larger than $\alpha = 0.02$ used by \citet{grzedzielski2017}. The solution roughly corresponds to the middle panel of Figure~\ref{fig:zestaw1} 

The source behavior, however, is not regular, the last outburst appeared earlier than expected and had higher amplitude than the remaining three. 
The optical V-band lightcurve reported by \citet{stanek2018} shows a small outburst lasting about one year, at around 2014, thus shorter by a factor of a few than the outbursts observed by \citet{alloin1986}. 

The duration of the outburst seems too short in comparison to the time separation. This is a characteristic property of the current version of the model, particularly for lower accretion rates, and large outburst amplitudes.

\subsubsection{NGC 4151}
For this source we assume mass from  \citet{woo2002} (log($M_{BH}/M\odot$) = 7.12) and bolometric luminosity 43.73 from \citet{2005ApJ...629...61K}, z = 0.003262 after NED. We estimate $\dot{m}$ as 0.027.
\citet{guo2014} suggest three possible periodicities for that source (P$_1$ = 4 $\pm$ 0.1, P$_2$ = 7.5 $\pm$ 0.3 and P$_3$ = 15.9 $\pm$ 0.3 yr). 
\citet{bon2012} also derive P = 15.9 yr for that source.  The same periodicity is also found in radial velocity curves of H$\alpha$ broad line.
Similar values were suggested by \citet{2018MNRAS.475.2051K} 
($\sim$ 5 and $\sim$ 8 years).
\citet{2007oknyanskij} suggest period about 15.6 years obtained using power spectrum. However, as \citet{czerny_power2003} shows, changes are not strictly periodic and possible period vary between 1/100 days and 1/10 years.

\begin{figure}
\centering
\includegraphics[width=0.45\textwidth]{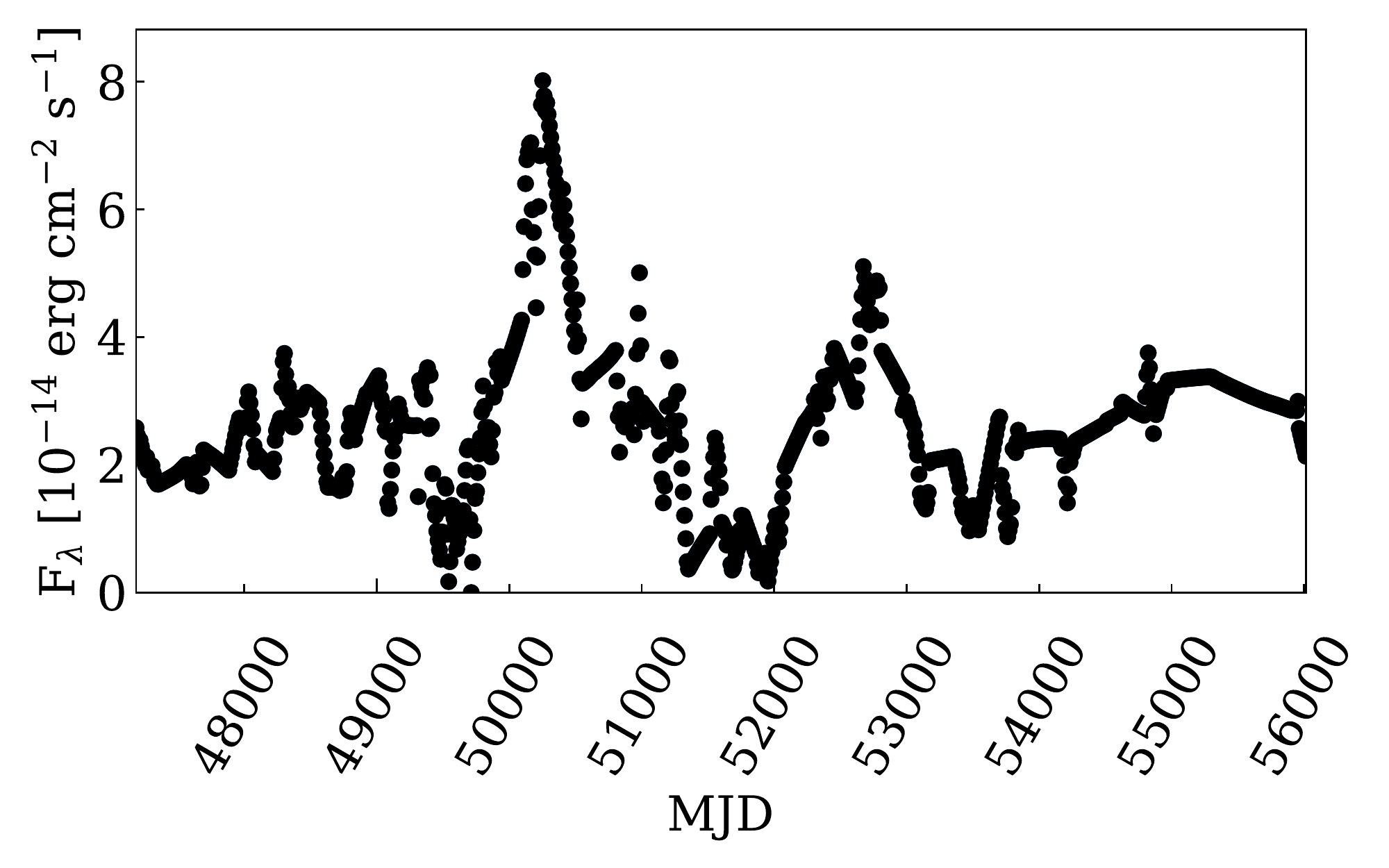}
\caption{The continuum flux evolution in NGC 4151 points.}

\label{fig:V_band_curve_4151}
\end{figure}

Photometry continuum flux data set includes data from: 
\citet{2006Bentz}, \citet{2008A&A...486...99S}, 
AGN Watch provided by \citet{1996ApJ...470..336K} and \citet{1997A&A...324..904M}.
To reduce the influence of the longest timescale systematic trend (which is probably due to different mechanism), we rebin the data. The result is shown in Figure~\ref{fig:V_band_curve_4151}. 


If we adopt the value of 10 years for a characteristic timescale in this source we require similar values of the remaining parameters as in the case of NGC 1566. For such parameters outbursts amplitudes are large, and outbursts rather short lasting.

\subsubsection{NGC 5548}
NGC 5548 is object with long-term and dense data coverage in various wavelengths \citep{2003chiang, 2015A&A...575A..22M, Mathur_2017}. Optical reverberation campaigns determined mass of its black hole \citep{peterson2004, 2015PASP..127...67B}.
This complex source is known by the changes of the BLR, which may not be linked with the only one physical origin.
 NGC 5548 showed the obscuration in X-ray and UV range \citet{2019kriss}, both the intrinsic continuum and the obscurer are variable \citet{2015A&A...579A..42D}. 
\citet{2019ApJ...877..119D} and references therein suggest that cloud shadowing should be considered as an appropriate explanation for variability in observation.

\begin{figure}
\centering
\includegraphics[width=0.45\textwidth]{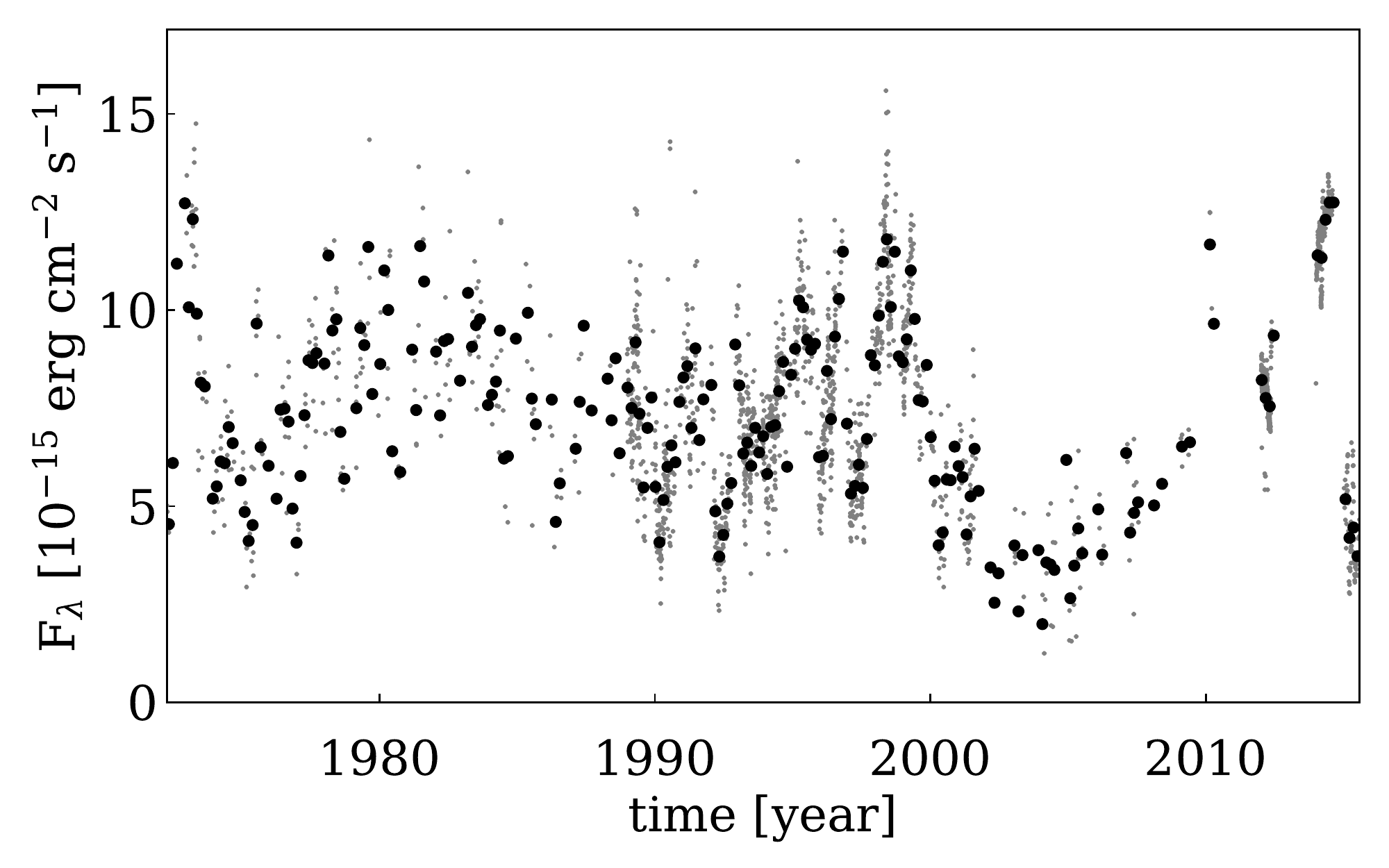}
\caption{The continuum flux evolution in NGC 5548. Gray points represent observational data, black points represent data rebinned to 300 bins in total.}
\label{fig:5548_data}
\end{figure}

We assume physical parameters for this source as follows:
$L_{BOL}$ = 44.45 from \citet{2016A&A...587A.129E}, M = $8.71^{+3.21}_{-2.61} \times 10^7 M_{\odot}$ from \citet{2016lu}.
\citet{2018MNRAS.475.2051K} suggest for that source period 13.3 $\pm$ 2.26 yr,
accretion rate 0.01 from \citet{papadakis2019}.
\citet{bon2016} suggested slightly longer period of $\sim$ 5700 days.
The continuum flux data set for NGC 5548 includes data from \citet{bon2016} is shown in Figure \ref{fig:5548_data}.  

We decided to model outbursts with period around 13 years. In that case, for the adopted mass and accretion rate as described above we can find the proper representation of the outbursts assuming much higher viscosity and somewhat smaller radius since the timescale is similar than in the two previous sources while the black hole mass is an order of magnitude higher. 

It is interesting to note that the location of the unstable zone in this source, at a distance of 0.23 light days from the center is nicely consistent with the location of the obscurer (below 0.5 light days) discussed by 
\citet{2019ApJ...882L..30D}.

\subsubsection{GSN 069}

\begin{figure}
\centering
\includegraphics[width=0.45\textwidth]{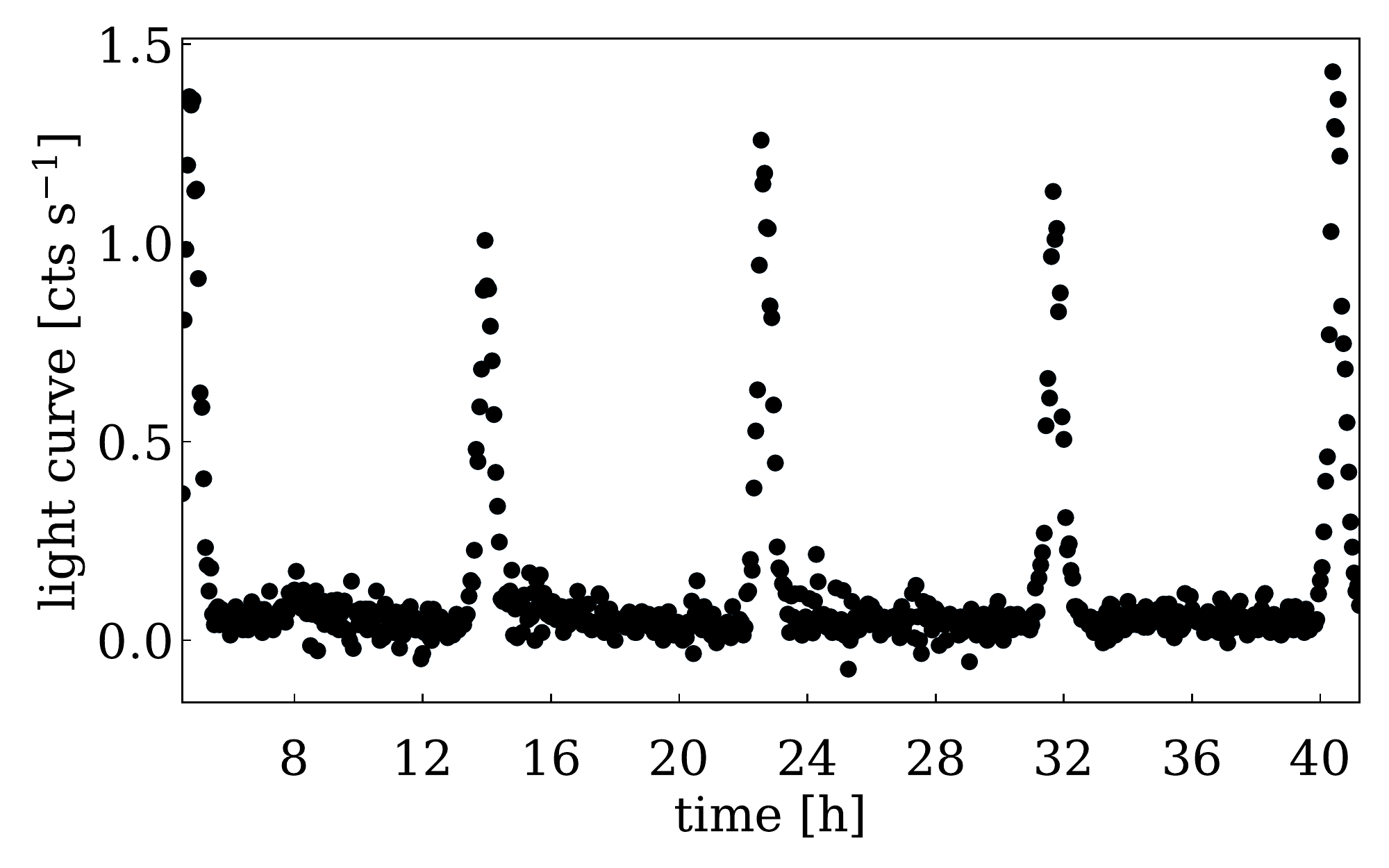}
\caption{GSN 069 disk contribution in 0.2-2 keV from \citet{Miniutti2019}}


\label{fig:gsn}
\end{figure}

This relatively low mass Seyfert galaxy ($M_{BH} = 4.5 \times 10^5 M_{\odot}$, \citealt{Miniutti2019}) was inactive when measured by ROSAT. In 2010 it showed a spectacular rise in the nuclear luminosity, followed by a slow decay. During the late decay phase, in December 2018, the source showed spectacular rapid large amplitude oscillations with the period of roughly 9 hours \citep{Miniutti2019}. The behavior was still observed in February 2019. The nature of these Quasi-Periodic Eruptions (QPE) is not clear but the spectral changes strongly suggest the coupling with the corona formation and likely the coronal inflow. The outbursts are shown in Figure~\ref{fig:gsn}.

We represented the variations in the disk luminosity using our toy model. 
We assumed the black hole mass of  $4.5 \times 10^5 M_{\odot}$, after \citet{Miniutti2019}, and we adjusted the remaining parameters to reproduce the timescale. QPE time separation can be indeed reproduced, although it requires small radius and large value of the viscosity coefficient. The external accretion rate favored by our model (0.013 in Eddington units) is much lower than the bolometric luminosity 0.46 estimated by \citet{Miniutti2019}. The low accretion rate was implied by the instability zone present very close to the black hole. Clearly the current toy model does not describe yet the source behavior, and most likely the source performed just disk/corona pulsations, as suggested by \citet{Miniutti2019}. If so, more complex model with two-phase disk/corona medium is needed to represent well this source.

\begin{table*}[]
\centering
\caption{Information about sources: Name of the source, redshift, luminosity distance [Mpc], assuming the cosmology: $H_O = 67$km s$^{-1}$, $\Omega_m = 0.32$,$\Omega_L = 0.68 $ \citep{2014planck}, black hole mass, bolometric luminosity, time span for the optical data coverage, the amplitude expected from the stochastic behavior of AGN, the observed amplitude from line or continuum.}
\begin{tabular}{lllllllll}
Name & redshift &d$_{L}$(z)   & log(M$_{BH}$) & L$_{BOL}$   & data coverage       & $\sigma$ exp & \begin{tabular}[c]{@{}l@{}}$\sigma$ obs \\ H$\beta$ \end{tabular} & \begin{tabular}[c]{@{}l@{}}$\sigma$ obs \\ cont\end{tabular} \\ \hline
\hline
NGC 1566 &0.005017 & 22.5 & 6.92    & $2.5 \times 10^{42}$ & 1972-1987 & 0.18  & 0.73                                                          & -                                                         \\
NGC 4151 &0.003319 & 14.9 & 7.12     & $7\times 10^{43}$   & 1988-2012 &  0.27     & -                                                             & 0.36                                                      \\
NGC 5548 &0.017175 & 77.8 & 7.94    & $2.8 \times 10^{44}$  & 1972-2015 &   0.20    & 0.40                                                          & 0.33 \\
GSN 069 & 0.018 & 81.6  &  5.65   & $2.7 \times 10^{43}$  & January 16/17, 2019   &   0.21     &                        -                &  1.85  \\ 
          & &  &    & & (130 [ks]) &      &  &    \\                   \hline                        
\end{tabular}
\label{table:observations}
\end{table*}


\begin{table*}[]
\centering
\caption{Summary of the results for each object. Name of the source, inner radius in $R_{Schw}$, the thickness of the unstable zone, viscosity parameter, period in years (except of the last object) and amplitude. Black hole mass used in the model was taken from Table~\ref{table:observations}}.
\begin{tabular}{lllllll}
Name     &  R$_{in}$ & $\Delta_R$ & $\alpha$ & $\dot{m}$ & period & amplitude \\ \hline
\hline
NGC 1566        & 25     & 0.002    & 0.04   & 0.015   & 5 yr  & 62.05     \\
NGC 4151        & 26     & 0.006    & 0.05   & 0.027   & 10 yr  & 341.3     \\
NGC 5548       & 20     & 0.003    & 0.20   & 0.0055  & 13 yr & 4892.36   \\
GSN 069         & 5      & 0.03     & 0.25   & 0.013   & 0.387 day & 3.95 \\ \hline
\end{tabular}
\label{table:fits-parameters}
\end{table*}

\section{Discussion} \label{sec:discussion}

The Changing-Look behavior - rapid large amplitude changes in active galaxies - is reported now with increasing frequency but the mechanism is still unknown. Part of the phenomena can be actually related to tidal disruptions, particularly in the case of a single, long lasting large amplitude event. On the other hand, if multiple event takes lace in a single source, particuarly in a semi-regular form, TDE mechanism is ruled out. There is al least one source where we observe a combination of the two phenomena: a galaxy GSN 069. Classified as Seyfert 2 galaxy, GSN 069 was undetected in ROSAT All Sky Survey, but in 2010 the source showed a spectacular brightening, and when the dimming continued, the source showed very regular outbursts \citep{Miniutti2019}.

In this paper we concentrate of modelling the repeating semi-regular outbursts observed in the sources radiating at a few per cent of the Eddington ratio.
We propose that the radiation pressure instability operating in the narrow zone between the outer gas-dominated stable accretion disk and an inner hot ADAF flow may be responsible for repeating outbursts in some CL AGN, as NGC 1566. We show that the proposed mechanism can lead to outbursts on a timescale much shorter than the usual viscous timescale in a cold disk. 

For this mechanism to operate we require that the mean accretion rate in the source is relatively low so the inner ADAF flow extends to a radius which is not much smaller than the radius where the radiation pressure in a standard Keplerian disk dominates. The amplitudes of the outbursts in the optical band are large due to the irradiation of the outer disk by the enhanced inner hot flow.

The generic prediction of the model is that the spectrum of the nucleus should become much bluer during outburst, and the outbursts in X-ray band should have comparable or larger amplitude. On the other hand, the current model does not  give the outburst shape consistent with the observational data. Our toy model relies  only on the viscous timescale in the unstable ring close to ADAF, without addressing the full complexity of the standard disk/ADAF transition. 

The model predicts strong semi-periodic variations in the emitted flux for sources at a few per cent of the Eddington ratio but these changes may, or may not be revealed in the properties of the BLR. If the variability timescale is long in comparison to the time required for the adjustment of the BLR structure to the change of the nuclear emission then the BLR will follow the changes in the nucleus in a quasi-stationary way, and we should see the classical Changing-Look AGN phenomenon. On the other hand, if the nuclear changes are lasting too shortly, then the BLR may not fully adjust. For example, in sources like GSN 069 the eruption lasts about an hour while the distance to the BLR is likely a few hours, like in another low mass Seyfert galaxy NGC 4395 \citep{peterson2005}. What is more, the light travel time describes just a change in irradiation while BLR structure adjusts more slowly \citep[see e.g.][]{hryniewicz2010}. 

\citet{ross2018} considered a possibility that the behavior of the quasar J1100-0053 is related to instability in a cold disk/ADAF  transition zone but argues against it since in other objects (e.g. NGC 1097) the transition zone is stable. Indeed, the position of the transition radius determined by balancing the cold disk evaporation rate and the inner hot flow depends on the global accretion rate \citep[e.g.][]{rozanska00,spruit2002,taam2012} seems rather stable. Our solution to the problem comes from introducing the radiation pressure instability. It also implies that for lower Eddington ratio objects the instability would not operate while for higher Eddington objects this mechanism would lead to outbursts of much larger part of the disk and it will operate on a timescale of thousands of years, as typically predicted for the radiation pressure instability \citep{janiuk2002,czerny2009,wu2016,grzedzielski2017}. However, if the evolution includes the time-dependent coronal flow \citep[e.g.][]{2007janiuk} or time-dependent vertical stratification of the disk into cold standard disk and the warm corona \citep[e.g.][]{2003corona,2020gronkiewicz,2020pop}, the timescales will be strongly affected due to quadratic dependence of the viscous timescale on ratio of the local medium geometrical thickness to the local radius.

In most cases the outbursts we model are not clearly quasi-periodic (the behavior of the source GSN 069 discovered by \citealt{Miniutti2019} is a nice exception) so there is a danger that we try to model the source behavior using a dedicated mechanism while in reality all AGN show a stochastic variability, and this stochastic variability may lead sometimes, with certain statistical probability, to a behavior which looks like quasi-periodic \citep{vaughan2016}. However, stochastic variability has a well defined power spectrum shape and normalization, both in X-rays \citep{mchardy2004} and in the optical band \citep[e.g.][]{czerny1999,czerny_power2003}.  Thus the amplitude for a given time span is limited. For the studied sources we thus report the observed variability amplitude for a given time period and we compare it from the amplitude expected from the stochastic behavior of AGN. For NGC 1566, NGC 4151 and NGC 5548 this stochastic amplitude was predicted assuming the power spectrum from the recent work of \citet{2010MNRAS.403..605B}, including the scaling by a factor 100 between the optical and the X-ray power spectrum, and the break in the X-ray spectrum for each source was estimated following \citet{2006Natur.444..730M}. Knowing the optical power spectrum we could predict the source variance expected from the standard stochastic variability. For GSN 069 we estimated the expected X-ray variability from the typical X-ray variability level of AGN \citet{2002MNRAS.332..231U} by averaging the provided values of $\sigma$ for four sources for timescales of order of $10^6$ s. The dispersion in those values was small, and we took the timescales longer than the length of the used GSN 069 lightcurve since the level of X-ray variability at a given timescale scales with the black hole mass \citep[e.g.][]{nikolajuk2004}. All values are reported in Table~\ref{table:observations}. We see that the observed dispersion is much larger than expected from the stochastic variations. Therefore, invoking a separate mechanism to explain this phenomenon is justified.

The presented toy model is still too simplistic to account quantitatively for the observed outbursts. The comparison shows that the duration of the outburst in some models is far too short in comparison with the rising phase while is some cases (GSN 069) they are too long. This is partially because the model does not account properly for the evaporation mechanism of the zone. In a realistic model Equation~\ref{eq:evap} should be replaced with the physically motivated equation containing the additional timescale for the process. However, this is not simple. The spectral changes observed in GSN 069 during outbursts \citep{Miniutti2019} suggest that a comptonizing corona forms above the disk, and it may be that the real mechanism is actually a two-step mechanism, with corona formation as a stage one, and then the corona inflow as a stage two, finishing outbursts. Thus the future model should have both radial and vertical stratification, perhaps actually a full 2-D since the height of the zone is comparable to its radial extension, and it should include full time-dependence of the outer disk since the irradiation would couple to the stability properties. However, such a model is far beyond the aim of the current project.

The second equally important aspect is the time delay in the signal propagation. The current model assumes that the change in the irradiation patters happens without any time delay while actually the outer parts of the disk react with significant time delay of days, as is well known from reverberation studies of AGN continua (e.g. \citealt{collier1998,sergeev2005,cackett2007,edelson2015,cackett2020}, and the references therein). The response of the emission lines is delayed even more strongly as showed by numerous campaigns (e.g. \citealt{liutyi1977,collier1998,kaspi2000,peterson2004,grier2017,du2018}), thus the final model has to include these effects, particularly in the case of relatively fast variations in the source.

\section*{Acknowledgements}
We thank Alex Markowitz for helpful discussion and Giovanni Miniutti for providing the data for the source GSN 069 and very helpful comments to the manuscript.
The project was partially supported by National Science Centre, Poland, grant No. 2017/26/A/ST9/00756 (Maestro 9),  and by the MNiSW grant DIR/WK/2018/12. Part of the work was done when BC was supported by a Durham Senior Research Fellowship COFUNDed between Durham
University and the European Union under grant agreement number 609412.
E.B. and N.B. acknowledge the support of Serbian Ministry of Education,
Science and Technological Development, contract number 451-03-68/2020/14/20002.

\section*{Software}
\citet{r}
%
%

\bibliographystyle{aa}
\bibliography{CL_v10_26_06}

\end{document}